\documentstyle[12pt,doublespace]{article}

\newcommand{\raf}[1]{(\ref{#1})}
\newcommand{\kv}{\mbox{${ \bf k}$}}                    
\newcommand{\qv}{\mbox{${ \bf q}$}}                    
\newcommand{\bitem}{\begin{itemize}}
\newcommand{\eitem}{\end{itemize}}
\newcommand{\kp}{\mbox{${ \bf k}_{+}$}}                
\newcommand{\km}{\mbox{${ \bf k}_{-}$}}                
\newcommand{\kw}{\mbox{$({\bf k},\omega)$}}
\newcommand{\rt}{\mbox{$({\bf r},t)$}}
\newcommand{\kt}{\mbox{$({\bf k},t)$}}

\newcommand{\rv}{\mbox{${\bf r}$}}

\newcommand{\mb}{\begin{equation}}                   

\newcommand{\men}{\end{equation}}                    
\newcommand{\pri}{\prime}

\newcommand{\w}{\mbox{$\omega$}}                     
\newcommand{\bea}{\begin{eqnarray}}
\newcommand{\nea}{\end{eqnarray}}
\newcommand{\pmj}{\mbox{$\phi_{m}$}}
\newcounter{fignum}
\begin{document}
\def\singlespacing{\baselineskip=12pt}
\def\doublespacing{\baselineskip=24pt}
\doublespacing
\mbox{}
\vspace{1cm}
\begin{center}
{\bf GENERALIZATIONS OF THE KPZ EQUATION} \\
\vspace{0.5cm}
{\it J. P. Doherty, M. A. Moore, J. M. Kim $^\ast$,
 A. J. Bray } \\
\vspace{0.5cm}
 Department of Theoretical Physics \\
University of Manchester \\ Manchester M13 9PL, England \\
\end{center}
\begin{abstract}
We generalize the KPZ equation to an O(3) $N=2j+1$ component
model. In the limit $N \to \infty$ we show that the mode coupling equations
become exact. Solving these approximately
we find that the dynamic exponent
$z$ increases from $3/2$ for $d=1$ to $2$ at the dimension $d\approx3.6$.
For $d=1$ it can be shown analytically that $z=3/2$ for all $j$. The
case $j=2$ for $d=2$ is investigated by numerical integration of the KPZ
equation.
\end{abstract}
\vspace{3in}
\newpage

Many growth models have been studied in recent
years both
analytically and numerically. A widely used description of
the broad scale features of such growing
surfaces is a non-linear Langevin
type equation proposed by Kardar,Parisi and Zhang (KPZ) \cite{KPZ} :
\mb
\frac{\partial}{\partial t}
\phi({ \bf r},t) =\nu\nabla^{2}\phi + \lambda(\nabla \phi)^{2}
+\eta({ \bf r},t).
\label{eq:KPZ}
\men It appears to describe the surface profiles generated in the
Eden model \cite{roux}
and  by ballistic deposition \cite{ballistic}.
It can also be easily mapped onto the directed polymer problem in a random
potential \cite{polymer} .
 It describes the time evolution of a single valued height parameter
$\phi({ \bf r},t)$ (ie with no overhangs or voids) for a growth
 process on a $d$
dimensional substrate. The equation reflects the competition on mesoscopic
length scales between surface tension smoothing forces $\nu\nabla^2 \phi$,
 the tendency for the growth to occur preferentially in the direction
of the local normal to the surface, represented by the term
$\lambda (\nabla \phi)^2$,
 and the noise which is Gaussian, such that
\mb
\bigl\langle\eta({\bf r},t)\eta({ \bf r}^\prime,t^\prime) \bigr\rangle
= 2D
\delta^d({ \bf r}-{ \bf r}^\prime) \delta(t-t^\prime) .
\label{eq:gnoise}
\men

The objective is to characterize the form of the surface. A step towards
this is to evaluate the two-point correlation function,
\mb
C\kw=
\frac{ \bigl\langle\phi\kw\phi^\ast(\kv^\prime,\w^\prime)\bigr\rangle}{
(2\pi)^{d+1}\delta^d(\kv-\kv^\prime) \delta(\w-\w^\prime)}
\label{eq:CC}
\men
and the response function in frequency space,
\mb
G\kw=
\frac{1}{
\delta^d({\bf k}-{\bf k}^\prime)\delta(\w-\w^\prime) }
\biggl\langle\frac{\partial \phi\kw}{\partial \eta({\bf
k}^\prime,\w^\prime)}\biggr\rangle .
\label{eq:GG}
\men
There are two distinct regimes in the scaling limit for $d>2$.
 In the weak coupling limit, i.e. for $\lambda<\lambda_c$,
 the behavior is governed by the $\lambda=0$ fixed point.
 In the strong coupling regime, $\lambda>\lambda_c$,
due to the non-linearity the correlation and response functions
take on the non-trivial scaling forms,
\bea
C\kw &=&  \frac{1}{k^{2\chi+d+z}}f_c\!\left(\frac{\w}{k^z}\right)
\nonumber \\
G\kw &=&  \frac{1}{k^z} f_g\!\left(\frac{\w}{k^z}\right) ,
\label{eq:scaling}
\nea
where $\chi$ is related to $z$  via the scaling relation $\chi+z=2$. In the
weak coupling regime $z=2$ and is independent of $d$. For $d\leq 2$ only the
strong coupling regime exists \cite{two,FNS}.
 Knowledge of the exponent $z$ as a function
of the dimensionality $d$, and the scaling
functions, would mean the system was essentially understood.

Some workers \cite{derrida,halpin,feigelman}
believe that above some finite
critical dimension $d_c$, $z=2$ for both the strong and weak
coupling regimes.
 This is in opposition to numerical work \cite{ala} which
suggests that the upper critical dimension is infinite,
although crossover effects might mask the true value of $z$ for the small
systems studied in higher dimensions.

A major theoretical difficulty is that in the strong coupling regime
the perturbation series in $\lambda$ about $\lambda=0$
cannot be summed self-consistently
 in terms of just response and correlation functions
because of vertex correction graphs
which renormalize the non-linearity as shown in {Fig. 1}.
 (The perturbation formalism we are using is that of Ref. \cite{FNS}.)
 However
as at each order in $\lambda$ graphs containing
vertex corrections scale in the same manner as graphs without vertex
corrections, it is tempting to ignore all the vertex correction graphs
 completely and sum the graphical series Fig. 2
 self-consistently , leading to the mode coupling equations
\cite{mcsource}.
 These equations are expected to give qualitatively correct
values for
the exponent $z$ yet they remain an uncontrolled approximation.
 Some modification
of the KPZ equation for which mode coupling was exact would place the
 approximation on a surer footing.

A widely used method in the study of critical phenomena is to increase the
number of components of the field to $N$, where it is often found that the
new model is exactly solvable
in the infinite  component limit and that
a systematic
$1/N$ expansion may be developed.
 It is possible to perform such generalizations in many ways.
 Kraichnan and Chen \cite{krak} in their formulation
of the DIA equations for turbulence
took a model where the vertex factors had a random
amplitude $\pm1/N$.
 As in the turbulence problem considered by Mou and Weichman \cite{wei},
 we use the approach of Amit and Roginsky \cite{amit}
 and generalize the KPZ equation to an $N=2j+1$ component field which forms the
basis for an
 irreducible representation of the O(3) symmetry group,
\mb
\frac{\partial }{\partial t}  \pmj\rt  =
\nu\nabla^{2}\pmj +
\frac{\lambda}{\sqrt{N}} \sum_{n,l}
 A^{l,n}_{m} \nabla\phi_{l} \nabla\phi_{n}+\eta_{m}({ \bf r},t).
\men
 The non-linear term in the KPZ equation represents the coupling of two fields
 to produce a field of index $m$,\,
 $m=-j,-j+1,\ldots,+j$. The noise has the form,
\mb
\langle \eta_m\rt \eta_{m^{\pri}}^{\ast}
({\bf r^\pri},t^\pri) \rangle = 2D \delta_{m,m^{\pri}}
\delta^d({\bf r}-{\bf r}^{\pri})\delta(t-t^{\pri})
\men
with $\eta_m^{\ast}=(-1)^m \eta_{-m}$.

 For the form of the generalized KPZ equation to be independent of
the representation it must be unchanged by the transformation
$ \phi_{m} \to \phi_{m^\pri}={\cal R}^{m}_{m^\pri}(u) \phi_{m}$
 given $u\in $ O(3) , ${\cal R}_{m^\pri}^{m}$ being
NxN matrices which form  an irreducible unitary
representation of O(3). The coupling tensor
$A_{m}^{n,l}$ must satisfy
\mb
A_{m^\pri}^{n^\pri,l^\pri}={\cal R}_{m^{\prime}}^{m}(u)\,
{\cal R}_{n}^{n^\pri}(u)\,  {\cal R}_{l}^{l^\pri}(u)\, A_{m}^{n,l}.
\men
The tensor $A_m^{l,n}$
is therefore the Clebsch-Gordan coefficient $\langle j,n;j,l |j,m \rangle $.
 The angular momentum $j$ must be an even integer
for non-zero $A^{l,n}_{m}$. The case $j=0$ ($N=1$) is the
 scalar KPZ equation \raf{eq:KPZ}.

We shall now consider the large $j$ limit.
 The structure of the graphs is the same as that of the $\phi^3$ field theory
studied by Amit and Roginsky \cite{amit}. Taking over their
results regarding the asymptotic form of Clebsch-Gordan coefficients
it can be seen that graphs with one or more vertex corrections
have magnitude $1/N$ or less \cite{power}
relative to other graphs with no vertex corrections
of the same order in $\lambda$.
 Thus all the vertex corrections are negligible
as $N\to\infty$ and the graphs for
the correlation and response functions may be summed self consistently
by performing the summations over  the component labels using the
orthogonality relation ,
\mb
\sum_{n,l}\langle j,l;j,n | j, m \rangle \langle j,l;j,n | j,m^\pri \rangle
= \delta_{m,m^\pri}.
\men

The correlation function
$C\kw=\langle \phi_m\kw \phi^\ast_n\kw\rangle$ and the response function
related to $\langle \partial \phi_m \kw / \partial \eta_n \kw \rangle$
vanish unless $n=m$ , and for $n=m$ they are $m$ independent  as a consequence
of the overall rotational invariance of the equations.

The mode coupling (MC)\cite{mcsource} equations can be written as,
\bea
C(\kv,\w)&=& 2D|G\kw|^2 + \nonumber \\
&& \frac{2\lambda^2}{(2\pi)^{d+1}}|G\kw|^2
 \int^{+\infty}_{-\infty}d\mu
\int^{+\infty}_{-\infty}d^d{\bf q}\,\,
(\kp. \km)^2 C(\kp,\w_{+}) C(\km,\w_{-})
\label{eq:MCC}
\\
G^{-1}(\kv,\w) &=& G_0^{-1}\kw + \nonumber \\
&& \frac{4\lambda^2}{(2\pi)^{d+1}}
 \int^{+\infty}_{-\infty}d\mu
\int^{+\infty}_{-\infty} d^d{\bf q}\,\,
(\kp. \km)(\kp. \kv) C(\kp,\w_{+}) G(\km,\w_{-})
\label{eq:MCG}
\nea
where $\kv_{\pm}=\frac{\kv}{2}\pm \qv$ and $\w_{\pm}=\frac{\w}{2}\pm \mu$.
The first term in Eq. \raf{eq:MCC} can be dropped in the scaling limit provided
$z<(d+4)/3$. We find this condition is satisfied by our solution.
 In Eq. \raf{eq:MCG}, $G_0^{-1}\kw=\nu k^2 -i\w$, but only the term
$-i\w$ is relevant in the scaling limit. Inserting the scaling forms for
 the response and correlation function into
the mode coupling (\ref{eq:MCC},\ref{eq:MCG})
 equations yields the scaling relation $\chi +z=2$.
 To calculate the exponent $z$ from
the MC equations we need to input the exact scaling forms for the response and
correlation functions in the right hand side of Eq. (\ref{eq:MCC},\ref{eq:MCG})
where
consistency will only be achieved for the exact value of $z$.
 A previous attempt by Bouchaud and Cates
 \cite{cates} to solve the MC equations (approximately) for
 the exponent $z$ by this consistency requirement took the response
function in $\kv,t$ space to be a simple exponential decay ,
\mb
G\kt = \exp\left(-k^z t\right)\,\theta(t)
\men
and the correlation function to be,
\mb
C\kw= {\cal D}k^{3z-d-4} |G\kw|^2,
\men
where ${\cal D}$ is a constant.
 We shall use instead as the starting point of our calculation the work
of Hwa and Frey \cite{hwa} in $d=1$ on the form of the scaling function.
 By solving the mode coupling equations numerically they showed that
\mb
G(\kv,t)\approx \exp(-\alpha \kv^2 t^{2/z})\,\theta(t)
\label{eq:G}
\men
was a good approximation for the response function \cite{hwa2}.
 The parameter $\alpha$ is dependent on the units used and may be scaled
out by the transformation $\kv \to \kv/\sqrt{\alpha}$.
 Since in $d=1$ and only in this dimension
the correlation and response functions satisfy a fluctuation
dissipation theorem (FDT)\cite{haake},
\mb
C(\kv,w) =\frac{1}{k^2}(G\kw+G^{\ast}\kw),
\label{eq:PDFT2}
\men
we take the form of the correlation function to be
\mb
C(\kv,t)=\frac{B}{k^{d+4-2z}}
\exp\left(- \kv^2 |t|^{2/z} \right)
\label{eq:C}
\men
where $B$ is some arbitrary constant that depends on $\lambda$.
 Note that the correlation function is an even function of $\w$ and that it
satisfies the FDT for $d=1$ and $z=3/2$.

 We shall assume that the form of the scaling functions do not vary strongly
 with the the substrate dimension $d$. We work away from $d=1$ using
the same choice for the scaling functions.
 The Gaussian form of the response and correlation functions
allows most of the integrations to be done analytically.
 Matching both sides of the fourier transformed versions of \raf{eq:MCC} and
\raf{eq:MCG} at $\w=0$
allows $z$ to be determined \cite{bray}.
 This matching procedure is an arbitrary choice out
of many schemes available but it is hoped that the exponent $z$ does not
depend significantly on the details of the matching.

 The form of the variation of $z$ with $d$  we find to be qualitatively
 similar to that of other self-consistent treatments
 \cite{cates,edwards}.
 We obtain the exact result $z=3/2$ for $d=1$.
 Our estimate for $z$ is $4/3$ as $d \to 0$. (This is the exact
 value for $z$ for any $j$ as $d\to 0$.)
 The numerical value of the exponent
was $z=1.662$ for $d=2$ which is close to the value found by
 Bouchaud and Cates \cite{cates}, $z=1.67$, but still some way off
the value obtained from simulations for $j=0$ \cite{forrest}
where $z\approx1.614$ .
 The exponent $z$ reaches its weak coupling value
$z=2$ at $d\approx3.6$ which is similar to the values found in Ref.
\cite{cates} $d_c\approx 3.75$ and significantly larger than
the value of Schwartz and Edwards \cite{edwards} $d_c\approx 3.25$.
 It should be noted that the form of the our scaling functions is
identical to that of Bouchaud and Cates for $z=2$ so
$d_c$, the upper critical dimension, should be the same in both approaches.
We do not understand this discrepancy.

The variation of $z$ with $d$ thus seems to be only weakly dependent
on the assumed form
of the scaling functions.
 The crucial question remains whether this apparent upper critical
dimension is an artifact of using inappropriate approximations for
 the scaling functions.
 We hope in future work to develop a systematic
procedure of matching the derivatives of the mode coupling
 equations at $\w=0$ . This
will provide better approximations for the form of the scaling functions
and more accurate estimates of $z$ away from $d=1$.
 It may also be possible to solve for the scaling function numerically (as Hwa
and
Frey did in $d=1$) for general dimension $d$.

The strong coupling value of the exponent
$z$ might be expected to vary continuously between $j=0$ (the scalar KPZ)
and the $j\to \infty$ limit (in which case mode coupling is exact).
 We have investigated the case of $j=2$ which is an intermediate value.
 For $j=2$ the field $\phi_m$ has 5 independent components. It is analogous
to the case of a traceless symmetric 3x3 tensor order parameter. We
calculated a rotationally invariant characteristic width of
the interface at equal times,
\mb
W(t)=
\left(
\frac{24\pi}{5}
\langle \sum_{m=-2}^{m=2} \phi_m(\rv,t)
\phi_m^\ast(\rv,t) \rangle \right)^{\frac{1}{2}}
\men
by numerically integrating the KPZ equation in $d=1$
and $d=2$ using a discrete grid method \cite{disc}.
 The width calculated at equal times for a finite size system of width
$L$ scales as,
\mb
W(L,t)= L^\chi f\!\left(\frac{t}{L^z}\right)
\men
 by analogy with the scalar KPZ equation \cite{equal}.
 For short times $W(L,t)\sim t^{\chi/z}$. We evaluated $\chi/z$
from the slope of a log-log plot of $W(L,t)$ versus $t$  calculated at
each time step $\delta t$ for a system with for $d=1$,
 $L=20,000,\,g=10$ ($g$ being the
magnitude of the effective coupling \,$g=2\lambda^2 D/\nu^3$)
and $\delta t=0.001$
to obtain,
$\chi/z=0.33\pm 0.02$.
 This value was found to be very robust for different $g$ and $\delta t$ and
is in agreement with the scalar KPZ equation in $d=1$.

The exponent $\chi$ is determined by the usual procedure of starting from
a flat surface of linear size $L$ and growing it until the fluctuations in
the height are saturated by finite size effects where $W(L)\sim L^\chi$.
 $W(L)$ was measured for system sizes $L=5,10,20,40,80,160$ in $d=1$.
 We plotted $\log(W(2L)^2-W(L)^2)$
versus $\log(L)$ to eliminate any constant correction to scaling. By
least squares fitting of the data we obtained $\chi=0.50\pm 0.01$.
 The value is the same as that of $\chi$ for the scalar KPZ equation.

We attempted to calculate the exponents by the same method for the case
$d=2$. The calculation was plagued by an instability
where the width $W(L,t)$ diverges after a certain period of time
$t_i$. This time was found to be only
weakly dependent on the time discretization
used. Smaller time steps $\delta t$ only
delayed the onset of the instability by a small amount.
 Furthermore $t_i$ was found to decrease as $\lambda$ was increased.
 For times less than $t_i$ a region exists in
which the $\log W(L,t)$ versus $\log t$ plot has a constant slope but this
slope was found to rise as $\lambda$ was increased. This may be due to some
crossover effect from weak to strong coupling.
 The maximum slopes are consistent with the exponent ratio $\chi/z\approx
0.16-0.2$ giving $z\approx 1.67-1.72$ .
 These difficulties are the same as those found for the scalar KPZ equation in
$d=3$ \cite{moser} , but the instability is more pronounced for $j=2$ making
reliable calculation of the exponents difficult. The instability may be due
to the discretization of time and space used in the numerical integration,
but further work is needed to substantiate this idea.

Using the methods of Huse, Henley and Fisher \cite{polymer} we have found
the stationary point solution in $d=1$ of the equivalent Fokker-Planck
equation for the $N$ component KPZ equation.
 From this we see that a fluctuation-dissipation theorem exists in $d=1$
for all even $j$, and that $z=3/2$.
 To observe the effect of varying
$j$ on the exponent $z$ the system has to be studied in $d=2$. As we
have seen, this presents a formidable numerical challenge.
 We speculate though that the generalized
KPZ equation for $j=2$ might actually describe a situation of physical
interest e.g. the time evolution of the orientational order
of a surface grown by aggregation of ellipsoids.

\newpage
\section*{ Figure Captions}

Fig. 1: An example of graphs which renormalize the value of the non-linear
vertex term in the perturbation series for the correlation function.
$\lambda$. In the large $N$ limit the contribution of these graphs is down by
a factor of $1/N$. The circle denotes the noise term $2D$.\\
Fig. 2: The graphical series neglecting vertex correction graphs
for the correlation $C\kw$ and response $G\kw$
respectively, which give the mode coupling equations.


\newpage
\begin{thebibliography}{99}
\bibitem [*] {address} Currently at the Department of Physics,
 University of Maryland,
 MD 20742, USA.
\bibitem {KPZ} M. Kardar, G. Parisi, and
Y. Zhang, Phys. Rev. Lett. {\bf 56}, 889
(1986).
\bibitem {roux} M. Eden, {\it Proc. 4$^{th}$ Berkeley Symposium
on Math. Stat. and Prob.}, Vol. 4 (Neyman F. ed., Univ. of California Press,
Berkeley), 223 (1961),
 S. Roux, A. Hansen, and E. L. Hinrichsen,
 J. Phys. A: Math. Gen. {\bf 24}, L295 (1991).
\bibitem {ballistic} M. J. Vold, {\it J. Colloid Sci.} {\bf 14}, 168 (1959),
J. Krug and H. Spohn, in {\it Solids far from Equilibria:
Growth, Morphology and Defects} C. Godriche ed. (Cambridge University Press,
 Cambridge 1991).
\bibitem {polymer} D. A. Huse, C. L. Henley, and D. S. Fisher, Phys. Rev. Lett.
{\bf 55}, 2924 (1985).
\bibitem {two} J. M. Kim, A. J. Bray, M. A. Moore, Phys. Rev. A {\bf 44},
4782 (1991).
\bibitem {FNS} D. Forster, D. R. Nelson and M. J. Stephen, Phys.
Rev. A {\bf 16}, 732 (1977).
\bibitem {derrida} J. Cook and B. Derrida, J. Phys. A {\bf 23}, 1523 (1990).
\bibitem {halpin} T. Halpin-Healey, Phys. Rev. Lett. {\bf 62}, 442 (1989).
\bibitem {feigelman} M. Feigelman, V. B. Geshkenbin, A. Larkin, and
 V. Vinokur, Phys. Rev. Lett. {\bf 63}, 2303 (1989).
\bibitem {ala} J. M. Kim and J. M. Kosterlitz, Phys. Rev. Lett. {\bf 62},
2289 (1989),
T. Ala-Nissila, T. Hjelt, and J. M. Kosterlitz,
 Europhys. Lett. {\bf
19}, 1 (1992).
\bibitem {mcsource} H. Van Beijeren, R. Kutner, and H. Spohn, Phys. Rev. Lett.
 {\bf 54}, 2026 (1985).
\bibitem {krak1} R. H. Kraichnan, J. Fluid Mech. {\bf 5}, 497 (1959).
\bibitem {krak} R. H. Kraichnan and S. Chen, Physica (Amsterdam) {\bf 37D}, 160
 (1989).
\bibitem {wei} C. Y. Mou and P. Weichman, Phys. Rev. Lett. {\bf 70}, 1101
(1993).
\bibitem {amit} D. Amit and D. V. I  Roginsky, J. Phys. A:Math. Gen.
 {\bf 12}, 689 (1979).
\bibitem{power} It is not known how the magnitude of the vertex correction
 graphs decreases as $N$ increases.
 Therefore the classes of graphs that would form the $1/N,1/N^2,\ldots$
 terms in a systematic expansion about $N\to \infty$ have not yet been
 identified, nor is it known whether even such an expansion exists.
\bibitem {cates} J. P. Bouchaud and M. E. Cates, Phys. Rev. E {\bf 47}, R1455
(1993) and unpublished.
\bibitem {edwards} M. Schwartz and S. F. Edwards, Europhys. Lett. {\bf 20}, 301
 (1992).
\bibitem {hwa} T. Hwa and E. Frey, Phys. Rev. A {\bf 44}, R7873 (1991).
\bibitem {hwa2} T. Hwa, E. Frey, and D. S. Fisher, unpublished.
\bibitem {haake} U. Dekker, F. Haake, Phys. Rev. A {\bf 11}, 2043 (1975).
\bibitem {bray} A.J. Bray, Phys. Rev. Lett. {\bf 32}, 1413 (1974).
\bibitem {forrest} L. H. Tang, B. M. Forrest, and D. E. Wolf, Phys. Rev. A
{\bf 45} , 7162 (1992).
\bibitem {disc} J.G. Amar and F. Family, Phys. Rev. A {\bf 41}, 3399 (1990).
\bibitem {equal} F. Family and T. Vicsek, J. Phys. A: Math. Gen.
  {\bf 18}, L75 (1985).
\bibitem {moser} K. Moser and J. Kertesz, Physica (Amsterdam)
{\bf 178A}, 215 (1991).
\end{thebibliography}
\end{document}